\documentclass[11pt]{article}
\usepackage{epsfig}
\usepackage{graphicx}
\usepackage{amssymb}
\usepackage{here}

\usepackage[figuresright]{rotating}

\newcommand{\AmS}{{\protect\the\textfont2
  A\kern-.1667em\lower.5ex\hbox{M}\kern-.125emS}}

\title{\Large{\bf Prospects of
probing new physics in bottomonium decays and spectroscopy}
\thanks{Research under grant FPA2002-00612 and GV-GRUPOS03/094.}}
\author{Miguel-Angel Sanchis-Lozano\thanks{Email:Miguel.Angel.Sanchis@uv.es}
\vspace{0.1cm}\\
\it Instituto de F\'{\i}sica
Corpuscular (IFIC) and Departamento de F\'{\i}sica Te\'orica \\
\it Centro Mixto Universidad de Valencia-CSIC \\
\it Dr. Moliner 50, E-46100 Burjassot, Valencia (Spain)}

\begin{document} 

\date{}
\maketitle

\begin{abstract}
A non-standard light
CP-odd Higgs boson could induce a slight (but observable) lepton 
universality breaking in Upsilon leptonic decays. Moreover,  
mixing between such a pseudoscalar Higgs boson and  $\eta_b$ states might
shift their mass levels, thereby modifying the
values of the $M_{\Upsilon(nS)}-M_{\eta_b(nS)}$ hyperfine splittings predicted 
in the standard model. Besides, $\eta_b$ resonances could  
be broader than expected with potentially 
negative consequences for discovery
in both $e^+e^-$ and hadron colliders. A scenario with a
CP violating Higgs sector is also considered. Finally, 
further strategies to search for a light Higgs particle 
in bottomonium decays are outlined.
\end{abstract}

\vspace{-13.1cm}
\begin{flushright}
  IFIC/05-19\\
  FTUV-05-0326\\
  March 26, 2005\\
  hep-ph/0503266
\end{flushright} 
\vspace{10.2cm}
\begin{small}
PACS numbers: 14.80.Cp, 13.25.Gv, 14.80.-j \\
Keywords: Non-standard Higgs, New Physics, bottomonium leptonic decays, 
lepton universality
\end{small}


\section{Introduction}

The search for $\lq\lq$new'' physics (NP) beyond the 
standard model (SM) has become one of the hottest topics
of the current decade. In most extensions of the SM,
new scalar (CP-even) and pseudoscalar (CP-odd) states 
appear in the physical spectrum. While the masses of these particles 
should be typically of the same order as the weak scale, if the theory
possesses a global symmetry its spontaneous breakdown gives rise to a
massless Goldstone boson, the $\lq\lq$axion'', originally 
introduced in the framework
of a two-Higgs doublet model (2HDM) \cite{gunion} to solve the  
strong CP problem. However, such an axial U(1) symmetry is anomalous and the 
pseudoscalar acquires a (quite low) mass ruled out experimentally.
On the other hand, if the global symmetry is explicitly (but slightly) broken, 
one expects a pseudo-Nambu-Goldstone boson in the theory which, for a 
range of model parameters, still can be significantly lighter than the 
other scalars. 

In the next to minimal supersymmetric standard model (NMSSM),  
where a new singlet superfield is added to the Higgs sector to solve
the so-called $\mu$-problem
\cite{gunion}, the mass of the lightest CP-odd Higgs can 
be naturally small 
due to a global symmetry of the Higgs potential only softly broken by
trilinear terms. This model has received considerable attention
and the associated
phenomenology should be examined with great care in
different experimental environments \cite{Hiller:2004ii}. 
For example,
it is likely that a SM-like Higgs boson would decay into two
(possibly) much lighter pseudoscalar Higgses presenting 
difficulties for detection at the LHC \cite{Dermisek:2005ar,Gunion:2004si}.

Also {\em Little Higgs} models can naturally
lead to the existence of light pseudoscalars (not absorbed as
longitudinal components of $Z'$ states) on account of
spontaneously broken $U(1)$ subgroups \cite{Kilian:2004pp}.
Moreover, there are other scenarios containing a light 
Higgs which could have escaped detection in the searches performed
at LEP-II \cite{Abbiendi:2004ww}, e.g. a MSSM Higgs sector
with explicit CP violation \cite{Carena:2002bb}. Another example is 
a minimal composite Higgs scenario
\cite{Dobrescu:2000yn} where the lower bound on the CP-odd scalar mass
is quite loose, as low as $\sim 100$ MeV (from astrophysical constraints).
On the other hand, it has been extensively argued in the literature
(see e.g. \cite{Cheung:2001hz,Czarnecki:2001pv}) 
that a light pseudoscalar Higgs should be required (in a
two-loop calculation) 
to account for the anomalous magnetic moment of the muon.

Long time ago, the authors of references \cite{haber79,ellis79}
pointed out the possibility of detecting a light 
Higgs particle in quarkonium decays. Recently, in a series of papers
\cite{Sanchis-Lozano:2004gh,Sanchis-Lozano:2003ha,Sanchis-Lozano:2002pm},
this investigation has been followed further by considering
a possible NP contribution
to the leptonic decays of $\Upsilon(nS)$ resonances (see figure 1) below
$B\bar{B}$ threshold via the decay mode: 
\begin{equation}
\Upsilon(nS)\ {\rightarrow}\ {\bf \gamma_s}\ 
A^0 ({\rightarrow}\ \ell^+\ell^-)\ ;\ 
\ell=e,\mu,\tau
\end{equation}
where $\gamma_s$ stands for a 
soft (undetected!) photon and $A^0$ denotes
a (real or virtual) non-standard light CP-odd Higgs boson.
Our later development is based upon the following keypoints: 
\begin{itemize}
\item Such a NP contribution would be unwittingly ascribed to the 
leptonic branching fraction: 
${\cal B}_{\ell\ell}={\cal B}[\Upsilon \to \ell^+ \ell^-]$
of the Upsilon. Notice that the experimental determination
of the leptonic width does include soft photons either from the initial- or 
final-state \cite{pdg}
\item A leptonic (squared) mass dependence of the decay width
(stemming from the Higgs contribution)
would lead to an $\lq\lq$apparent'' \footnote {In the sense
that lepton universality would be restored once
the Higgs contribution were taken into account} 
lepton universality breakdown. Actually, only in the tauonic channel
would this NP contribution significantly alter the measured branching
fraction (BF), as we 
shall later show. Let us also note that those $\Upsilon$ decays
breaking lepton universality within the SM
(like a direct $Z^0$-exchange annihilation 
or a two-photon (one-loop) annihilation of an intermediate $\eta_b$ state) 
are negligible 
\cite{Sanchis-Lozano:2003ha}

\end{itemize}

The electromagnetic decay width of the $\Upsilon(nS)$ resonance 
into a dilepton as a first approximation is given by \cite{royen}
\begin{equation}
{\Gamma}^{(em)}_{{\ell}{\ell}}= 
4\alpha^2 Q_b^2\ \frac{|R_n(0)|^2}{M_{\Upsilon}^2}\ {\times}\ 
K(x_{\ell})
\end{equation}
where $\alpha\ {\simeq}\ 1/137$ is the electromagnetic fine 
structure constant;
$m_{\Upsilon}$ denotes the mass of initial-state vector resonance, 
$R_n(0)$ its non-relativistic radial wave function at the origin;
$Q_b$ is the charge of the relevant (bottom) quark ($1/3$ in units of $e$);
$K(x_{\ell})=(1+2x_{\ell})(1-4x_{\ell})^{1/2}$
is a (smoothly) decreasing function of 
$x_{\ell}=m_{\ell}^2/M_{\Upsilon}^2$ with $m_{\ell}$ the lepton mass.

In order to check our conjecture by
assessing the relative importance of the postulated NP
contribution, we defined in \cite{Sanchis-Lozano:2004gh,Sanchis-Lozano:2004zd}
the ratio: 
\begin{equation}
{\cal R}_{\tau}=\frac{\Gamma[\Upsilon(nS)\to\gamma_s\ \tau^+\tau^-]}
{\Gamma^{(em)}_{ee}}=
\frac{{\cal B}_{\tau\tau}-\bar{\cal B}_{ee}}
{\bar{\cal B}_{ee}}
\end{equation}
where $\bar{\cal B}_{ee}=({\cal B}_{ee}+{\cal B}_{\mu\mu})/2$
stands for the mean BF
of the electronic and muonic modes of the $\Upsilon(nS)$. A
(statistically significant) non-null value of 
${\cal R}_{\tau}$ would imply the rejection
of lepton universality (predicting ${\cal R}_{\tau}=0$) and a strong argument
supporting the existence of a pseudoscalar Higgs boson mediating
the tauonic channel as shown in Eq.~(1).

\begin{figure}
\begin{center}
\includegraphics[width=14pc]{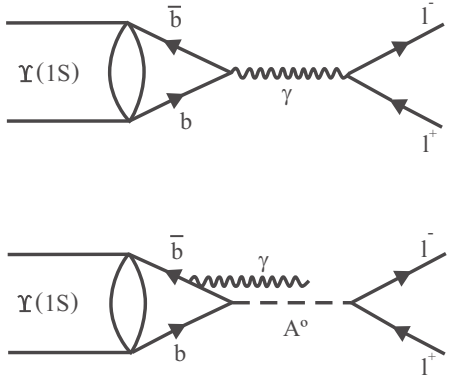}
\end{center}
\caption{
(a): Conventional electromagnetic annihilation of
the $\Upsilon(1S)$ resonance into 
a $\ell^+\ell^-$ pair. (b):
Non-standard Higgs-mediated annihilation subsequent to 
a (soft) photon emission either on the continuum or through
an intermediate $b\bar{b}$ bound state}
\end{figure}

Table 1 shows current experimental data (from \cite{pdg}) and
the corresponding ${\cal R}_{\tau}$ values for the $\Upsilon(1S)$ and
$\Upsilon(2S)$ resonances, the latter with a big error. Nevertheless,
we can conclude that those results don't preclude the possibility of a  
$\sim 10\%$ breaking of lepton universality (i.e. $R_{\tau} \sim 0.1$). 
Forthcoming data from CLEO on-going analisis
will definitely settle this point \footnote{New determinations of the muonic
BF of the $\Upsilon(1S)$, $\Upsilon(2S)$ and 
$\Upsilon(3S)$ resonances are available from CLEO (not used in Table 1)
but not the tauonic BF yet 
\cite{Danko:2004sb,Blusk:2004mm,Adams:2004xa}}, thus 
checking our conjecture.

For theoretical estimates we
will assume that fermions couple to the $A^0$ field
according to the interaction term
\[
{\cal L}_{int}^{\bar{f}f}\ =\ -\xi_f^{A^0}\ \frac{A^0}{v}
m_f\bar{f}(i\gamma_5)f 
\]
in the effective Lagrangian, with $v \simeq 246$ GeV; $\xi_f^{A^0}$ 
depends on
the fermion type, whose mass is denoted by $m_f$.
In what follows, we will focus on a
2HDM of type II \cite{gunion}: $\xi_f^{A^0}=\tan{\beta}$ for down-type
fermions where $\tan{\beta}$ stands for the ratio of two Higgs doublets
vacuum expectation values.  
Let su remark that $\xi_f^{A^0}=\cot{\beta}$ 
in the corresponding Yukawa coupling of up-type
fermions. Large values of $\tan{\beta}$ 
would imply a large coupling of the $A^0$ to the bottom quark but a
small coupling to the charm quark.
This fact has crucial phenomenological consequences in our proposal 
as a Higgs-mediated contribution would only affect 
bottomonium decays but not charmonium decays. Thus, in this work
we focus on $\Upsilon$ resonances to find out a possible signal of NP.

\begin{table*}[hbt]
\setlength{\tabcolsep}{0.4pc}
\caption{Measured leptonic branching fractions 
${\cal B}_{\ell\ell}$ (in $\%$) and error bars (summed in quadrature) of
$\Upsilon(1S)$ and $\Upsilon(2S)$ resonances 
(from \cite{pdg}); ${\cal R}_{\tau}$ is defined in Eq.~(3).}

\label{FACTORES}

\begin{center}
\begin{tabular}{ccccc}
\hline
channel: & $e^+e^-$ & $\mu^+\mu^-$ & $\tau^+\tau^-$ &  ${\cal R}_{\tau}$ \\
\hline
$\Upsilon(1S)$ & $2.38 \pm 0.11$ & $2.48 \pm 0.06$ & $2.67 \pm 0.16$ & 
$0.10 \pm 0.07$\\
\hline
$\Upsilon(2S)$ & $1.34 \pm 0.20$ & $1.31 \pm 0.21$ & $1.7 \pm 1.6$ & 
$0.28 \pm 1.21$ \\
\hline
\end{tabular}
\end{center}
\end{table*}

\section{Estimates according to a 2HDM(II)}

In this section we deal with different 
Higgs-mediated decay channels, either on the $\lq\lq$continuum''
i.e. without formation of bound 
$b\bar{b}$ states subsequent to
the photon emission in process (1), or via 
intermediate $\eta_b$ states. 
We will consider both off-shell and on-shell Higgs production according to 
whether the Higgs mass is greater or smaller than
the decaying Upsilon mass, respectively.

\subsection{Perturbative calculation without intermediate bound states}

Let us perform a 
perturbative calculation of the three body decay
$\Upsilon(nS) \to \gamma_s A^{0*}(\to \ell^+ \ell^-)$ on
the $\lq\lq$continuum'' by writing the
decay width as the integral over phase space
\begin{equation} 
\Gamma[\Upsilon(nS)\to\gamma\ \ell^+\ell^-]=
\frac{1}{32M_{\Upsilon}^3}\frac{1}{(2\pi)^3}\ \times\  
\int | {\cal A}(\Upsilon(nS)\to\gamma\ \ell^+\ell^-)|^2\ dm_{\ell\ell}^2\ 
dm_{\ell\gamma}^2 
\end{equation}
with
\begin{equation}  
| {\cal A}[\Upsilon(nS)\to\gamma\ \ell^+\ell^-]|^2 =
 \frac{64m_{\ell\ell}^2 m_b^2\alpha Q_b^2|R_n(0)|^2
\tan^4\beta}
{M_{\Upsilon}[m_{\ell\ell}^2-M_{A^0}^2]^2v^4}\ \times m_{\ell}^2
\end{equation}
Under the assumption that $M_{A^0} \gtrsim  M_{\Upsilon}$, i.e. no on-shell
production of the CP-odd Higgs boson is kinematically allowed, we
carry out the phase space integration in Eq.~(4);   
the leading term turns out to be
\begin{equation}
\Gamma[\Upsilon(nS)\to\gamma\ \ell^+\ell^-]\ \simeq\ 
\frac{\alpha|R_n(0)|^2}{144\pi^3v^4}
\biggl[\log{\biggl(\frac{M_{A^0}^2}{M_{A^0}^2-M_{\Upsilon}^2}\biggr)-1\biggr]}
\ \times\ m_{\ell}^2\ \ \ ;\ \ \ M_{A^0} > M_{\Upsilon}
\end{equation}
Only for the
tauonic mode would the NP contribution be noticeable
because of the Higgs coupling proportional to
the lepton mass, while the contribution to the
electronic and muonic modes is currently beyond experimental 
test (see Table 1). Thus one obtains for the ratio (3):
\begin{equation}
{\cal R}_{\tau}\ 
\simeq\ \frac{M_{\Upsilon}^2\tan{}^4\beta}{64 \alpha \pi^3v^4}
\biggl[\log{\biggl(\frac{M_{A^0}^2}{M_{A^0}^2-M_{\Upsilon}^2}\biggr)-1\biggr]}
\times\ m_{\tau}^2
\end{equation}

In order to get ${\cal R}_{\tau} \simeq 0.1$  (as suggested 
by current data shown in Table 1) from the continuum
setting $M_{A^0} \simeq 10$ GeV, rather large values of 
$\tan{\beta}$ are required, e.g. $\tan{\beta} \gtrsim 50$ for
$M_{A^0}-M_{\Upsilon}=0.25$ GeV, as concluded in 
\cite{Sanchis-Lozano:2003ha}.

\subsection{Intermediate bound states}

In Ref.~\cite{Sanchis-Lozano:2003ha} we used time-ordered
perturbation theory to incorporate the effect of intermediate 
$b\bar{b}$ bound states in the process (1). We found that the main contribution
should come from a $\eta_b$ state subsequent to an allowed 
dipole magnetic (M1) transition of the 
$\Upsilon$ vector resonance, i.e.  
\begin{equation}
\Upsilon{\rightarrow}\ {\bf \gamma_s}\ 
\eta_b (\rightarrow 
A^{0*} {\rightarrow}\ \ell^+\ell^-)\ ;\ 
\ell=e,\mu,\tau
\end{equation}
Thus, the total decay width 
can be factorized as \cite{Sanchis-Lozano:2003ha}
\begin{equation}
\Gamma[\Upsilon\to\gamma_s\,\ell^+\ell^-]\ =\ 
\Gamma^{\,M1}_{\,\Upsilon\to\gamma_s\eta_b}\ 
\times\ \frac{\Gamma[\eta_b\to\ell^+\ell^-]}{\Gamma_{\eta_b}}
\label{eq:factor1}
\end{equation}
where $\Gamma[\eta_b\to\ell^+\ell^-]$ and
$\Gamma_{\eta_b}$ denote the leptonic width and the total width
of the $\eta_b$ resonance respectively; 
$\Gamma^{\,M1}_{\,\Upsilon\to\gamma_s\eta_b}$ stands for
the M1 transition width. 

Dividing both sides of (9) by
the $\Upsilon$ total width, we get
the cascade decay formula
\[
{\cal B}[{\,\Upsilon\to\gamma_s\,\ell^+\ell^-}]\ =\ 
{\cal B}[{\,\Upsilon\to\gamma_s\eta_b}] \times
{\cal B}[\eta_b \to \ell^+ \ell^-] 
\]
The branching ratio for a magnetic dipole (M1) transition
between $\Upsilon(nS)$ and $\eta_b(nS)$ states
can be written in a non-relativistic approximation as 
\begin{equation}
{\cal B}[{\,\Upsilon\to\gamma_s\eta_b}] = 
\frac{\Gamma^{M1}_{\Upsilon{\rightarrow}\gamma_s\eta_b}}
{\Gamma_{\Upsilon}}\ {\simeq}\ 
\frac{1}{\Gamma_{\Upsilon}}\frac{4\alpha Q_b^2k^3}{3m_b^2} 
\label{eq:probability}
\end{equation}
where $k$ stands for the soft photon energy (approximately 
equal to the hyperfine 
splitting $M_{\Upsilon}-M_{\eta_b}$). Hindered M1 transitions
of the $\Upsilon(2S)$ and $\Upsilon(3S)$ resonances
into $\eta_b(2S)$ and
$\eta_b(1S)$ states should also be taken into account as
potential contributions to the process (8)
for such resonances.

The decay width of the $\eta_b$ into a dilepton mediated by
a $A^0$ boson reads in a 2HDM(II)
\[
\Gamma[\eta_b \to \ell^+\ell^-]\ =\  
\frac{3m_b^4m_{\ell}^2(1-4x_{\ell})^{1/2}
|R_n(0)|^2\tan{}^4\beta}
{2\pi^2(M_{\eta_b}^2-M_{A^0}^2)^2v^4}\ \simeq\  
\frac{3m_b^4m_{\ell}^2\tan{}^4\beta}{32\pi^2Q_b^2\alpha^2
{(1+2x_{\ell})\Delta}M^2v^4}\ {\times}\ 
\Gamma_{\ell\ell}^{(em)}
\]
where $\Delta M = | M_{A^0}-M_{\eta_b} |$ stands for the (absolute) 
mass difference
between the $\eta_b$ and $A^0$ states; $\Gamma_{\ell\ell}^{(em)}$
is given in Eq~(2). Note again that
only for the tauonic mode, would the NP contribution to the $\Upsilon$
leptonic decay be significant \footnote{Leaving aside
the unlikely case when the Upsilon and Higgs masses were very close
(within the MeV range)}.
Finally one gets 
\begin{equation} 
{\cal R}_{\tau} \simeq \biggl[\frac{m_b^2k^3\tan{}^4\beta}
{8\pi^2{\alpha}(1+2x_{\tau})\Gamma_{\Upsilon} v^4}\biggr] 
\times \frac{m_{\tau}^2}{{\Delta}M^2}
\end{equation}

For large $\tan{\beta}$ ($\gtrsim 35$) the
NP contribution would almost saturate the $\eta_b$ decay, i.e.
$\Gamma_{\eta_b} \simeq \Gamma[\eta_b\to\tau^+\tau^-]$; thus
${\cal B}[\eta_b \to \tau^+\tau^-] \simeq 1$ and consequently
\[ {\cal R}_{\tau} \simeq \frac{{\cal B}[\Upsilon\to\gamma_s\eta_b]}
{\bar{\cal B}_{ee}}\ \simeq\ 1-10\ \% \]
for $k=50-150$ MeV and $\Delta M = 0.25$ GeV. In fact, the quest for
a light Higgs particle would coincide with the search for $\eta_b$ states!

On the other hand, it is well known that higher Fock components beyond 
the heavy quark-antiquark pair can play
an important role in both production and decays 
of heavy quarkonium \cite{Bodwin:1994jh}. Thus, in 
\cite{Sanchis-Lozano:2003ha,Sanchis-Lozano:2002pm} 
we relied, as a factorization alternative to Eq.~(9), 
on the separation between 
long- and short-distance physics following the main lines of 
Non-Relativistic QCD \cite{Bodwin:1994jh} - albeit replacing a gluon 
by a photon in the usual Fock decomposition 
of hadronic bound states. Hence
we considered the existence in the Upsilon resonance of the $b\bar{b}$ pair 
in a spin-singlet and color-singlet state, i.e. a 
$|\eta_b^*+\gamma_s\rangle$ Fock component 
with probability ${\cal P}^{\Upsilon}(\eta_b^*\gamma_s) \simeq 10^{-4}$
\cite{Sanchis-Lozano:2003ha}.
Thus the total decay width can be factorized as
\begin{equation}
\Gamma[\Upsilon\to\gamma_s\tau^+\tau^-]\ =\ 
{\cal P}^{\Upsilon}(\eta_b^*\gamma_s) \times
\Gamma[\eta_b^*\to\tau^+\tau^-]
\end{equation}
In order to get a 10\% lepton universality breaking 
effect, a value of $\tan{\beta} \simeq 15$ is required setting again 
$\Delta M = 0.25$ GeV as a reference value. Those ranges of
$\tan{\beta}$ needed for larger values of $\Delta M$ can be found
in \cite{Sanchis-Lozano:2004gh}.

\subsection{On-shell Higgs production from $\Upsilon$ radiative decays}

Let us consider now a scenario where the Upsilon state lies
slightly above the lightest (CP-odd) Higgs state, i.e. $ M_{A^0}
\lesssim M_{\Upsilon}$. Then the decay
into an on-shell $A^0$ can proceed via the radiative process 
$\Upsilon \to \gamma + A^0$, whose width satisfies  the ratio \cite{wilczek77}
\[ \frac{\Gamma[\Upsilon \to \gamma A^0]}{\Gamma[\Upsilon \to \mu^+\mu^-]}
\simeq \frac{m_b^2\tan{}^2\beta}
{2\pi\alpha\ v^2}\biggl(1-\frac{M_{A_0}^2}{M_{\Upsilon}^2}\biggr) \ \ ;\ \ \ \
M_{A^0} < M_{\Upsilon}
\]

On the other hand, the decay width of a CP-odd Higgs boson into
a tauonic or a $c\bar{c}$ pair in the 2HDM(II)
can be obtained, respectively, from the expressions: \cite{Drees:1989du}
\begin{eqnarray}
\Gamma[A^0 \rightarrow \tau^+\tau^-] & \simeq & 
\frac{m_{\tau}^2\tan{}^2\beta}{8\pi v^2}\ M_{A^0}\ (1-4x_{\tau})^{1/2} \\
\Gamma[A^0 \rightarrow c\bar{c}] & \simeq & 
\frac{3m_c^2\cot{}^2\beta}{8\pi v^2}\ M_{A^0}\ (1-4x_c)^{1/2}
\end{eqnarray}
where $x_{\tau}=m_{\tau}^2/M_{A^0}^2$ and $x_c=m_c^2/M_{A^0}^2$.
Below open bottom production and above $\tau^+\tau^-$
threshold, the $A^0$ decay mode would be dominated by the tauonic channel
even for moderate $\tan{\beta}$. Therefore, the radiative decay should be
almost saturated by the channel:
\[ \Upsilon \to \gamma + A^0(\to \tau^+\tau^-) \] 
since ${\cal B}[A^0 \to \tau^+\tau^-] \approx 1$; then one may conclude that
\[ 
{\cal R}_{\tau} 
\simeq \frac{2M_{\Upsilon}^2\ \tan{}^2\beta}
{\pi\alpha\ v^2}\biggl(1-\frac{M_{A^0}^2}{M_{\Upsilon}^2}\biggr)
\]
Setting, e.g., $\tan{\beta}=15$, one gets ${\cal R}_{\tau} \simeq 10\%$.
In using a relativistic theory of the decay of
Upsilon into a Higgs boson plus photon in the mass range 7-9 GeV,
the ratio is substantially smaller (by an order-of-magnitude)
than that of the nonrelativistic calculation
presented above \cite{faldt:1988}. Then somewhat larger values of
$\tan{\beta}$ would be needed to yield ${\cal R}_{\tau}=0.1$.

Furthermore, also a light non-standard
CP-even Higgs boson (usually denoted as $h^0$) has not
been discarded by LEP searches \cite{Abbiendi:2004ww}. In such a case, 
the cascade decay 
\[ \Upsilon \to \gamma_s\chi_{b0}(\to h^0 \to \tau^+\tau^-) \]
could also ultimately contribute to enhance the tauonic decay modes of the
$\Upsilon(2S)$ and $\Upsilon(3S)$ resonances and
should not be overlooked. In fact, 
a discrimination between different sorts of Higgs bosons
(i.e. determining the CP quantum numbers) might be performed
by realizing whether the $\chi_{b0}$ resonances play a role
as intermediate states in the process (1).

The decay width of a $\chi_{b0}$ resonance into a tauonic pair
via a CP-even Higgs boson is \cite{haber79}
\begin{equation}
\Gamma[\chi_{b0} \to \tau\tau]=
\frac{27m_b^2m_{\tau}^2(1-4x_{\ell})^{3/2}|R'_n(0)|^2\tan{}^4\beta}
{8\pi^2(M_{\chi_{b0}}^2-M_{A^0}^2)^2v^4}
\end{equation}
where $R'_n(0)$ denotes the derivative of the 
$\chi_{b0}(nS)$ radial wave function at the origin,
yielding $\Gamma[\chi_{b0} \to \tau\tau] \simeq 20$ keV for
$|R'_n(0)|^2 \simeq 1.5$ GeV$^5$ \cite{eichten}, $\tan{\beta}=15$  
and $M_{\chi_{b0}}-M_{A^0}=0.25$ GeV as reference values.
In order to get a branching fraction estimate, let us normalize this 
width to 
\[ \Gamma_{\chi_{b0}} \approx \Gamma[\chi_{b0} \to gg]\ =\  
\frac{96\alpha_s^2|R_n'(0)|^2}{M_{\chi_{b0}}^4} \]
Setting $\alpha_s(M_{\Upsilon}) \simeq 0.15$ one gets
${\cal B}[\chi_{b0} \to \tau^+\tau^-] \simeq 6\%$. 
Now, since the radiative $\Upsilon$ decay rates into $\chi_{b0}$ are of order
$3-5\%$ \cite{pdg}, we can estimate for the combined branching ratio
\begin{equation}
{\cal B}[\Upsilon \to \gamma_s \tau^+\tau^-]\ =\ 
{\cal B}[\Upsilon \to \gamma_s \chi_{b0}]\ 
\times\ {\cal B}[\chi_{b0} \to \tau^+\tau^-]
\ \simeq\ 0.002-0.003
\end{equation}
yielding ${\cal R}_{\tau} \simeq 10\%$. Obviously, the
search for a light Higgs via this cascade decay channel 
does not apply to the $\Upsilon(1S)$ resonance.

\section{Possible spectroscopic consequences}

The existence of a light Higgs boson can have consequences
besides altering the tauonic decay rate of Upsilon resonances. Indeed, 
the mixing of the $A^0$ with a pseudoscalar resonance 
could modify the properties of both 
\cite{Drees:1989du,opal}. In particular, the 
resulting mass shift of the $\eta_b$ state might cause
a disagreement between the 
experimental determination of
the $M_{\Upsilon(nS)}-M_{\eta_b(nS)}$ hyperfine splittings  
and theoretical predictions based on quark potential
models, lattice NRQCD or pNRQCD \cite{Pineda:2004qq}. 
The masses of the mixed (physical) states in terms of
the unmixed ones (denoted as $A_0^0, \eta_{b0}$) are:
\cite{Drees:1989du}
\begin{equation}
M_{\eta_b,A^0}^2  = \frac{1}{2}(M_{A_0^0}^2
+M_{\eta_{b0}}^2) 
 \pm  \frac{1}{2}\biggr[\ (M_{A_0^0}^2
-M_{\eta_{b0}}^2)^2+4(\delta M^2)^2\ \biggl]^{1/2} 
\end{equation}
where $\delta M^2  \simeq  0.146 \times \tan{\beta}$ GeV$^2$.
For some mass intervals, the above formula simplifies to:
\begin{eqnarray}
M_{\eta_b,A^0} & \simeq & 
M_{\eta_{b0}}\ \mp\ \frac{\delta M^2}{2M_{\eta_{b0}}}\ \ ;\ \ 
0 < M_{A_0^0}^2-M_{\eta_{b0}}^2 << 2\ \delta M^2, \nonumber \\
M_{\eta_b,A^0} & \simeq & 
M_{\eta_{b0,A_0^0}}\ \mp\ \frac{(\delta M^2)^2}
{2M_{\eta_{b0}}(M_{A_0^0}^2-M_{\eta_{b0}}^2)}\ \ ;\ \ 
M_{A_0^0}^2-M_{\eta_{b0}}^2 >> 2\ \delta M^2 \nonumber
\end{eqnarray}
Setting $\tan{\beta}=20$ and  
$M_{\eta_{b0}} \simeq M_{A_0^0}=9.4$ GeV, as an illustrative example, one gets
$M_{\eta_b} \simeq 9.24$ GeV and
$M_{A^0} \simeq 9.56$ GeV yielding 
${\cal B}[\Upsilon(1S) \to \gamma \eta_b(1S)] \simeq 10^{-2}$. A caveat is 
thus in order: a 
quite large $M _{\Upsilon}-M_{\eta_b}$ difference
may lead to an unrealistic M1 transtion rate 
requiring smaller $\tan{\beta}$ values, in turn 
inconsistently implying a smaller
mass shift; hence no hyperfine splitting 
greater than $\sim 200$ MeV should be expected on these grounds.

On the other hand, broad $\eta_b$ states would be  
possible due to the NP contribution, notably 
for large $\tan{\beta}$ values. Thus one can speculate 
why no evidence of
hindered M1 radiative decays of higher Upsilon
resonances into $\eta_b(1S)$ and $\eta_b(2S)$ states  was found
in the search performed by CLEO \cite{Seth:2005pr,Artuso:2004fp,Blusk:2004mm}.
The corresponding signal peak (which should appear in the 
photon energy spectrum) could be considerably
smoothed - in addition to the spreading from the experimental
measurement - and thereby might not be significantly
distinguished from the background (arising
primarly from $\pi^0$'s). Of course, the 
matrix elements for hindered
transitions are expected to be small and difficult to 
predict as they are generated by relativistic and finite size
corrections. Nevertheless, most of the theoretical calculations
(see a compilation in Ref.\cite{godfrey})
are ruled out by CLEO results (at least) at $90\%$ CL,
though substancially lower rates are obtained in \cite{lahde} where
exchange currents play an essential role and therefore cannot be
currently excluded. 

Large widths of $\eta_b$ resonances would also bring negative effects for
their detection in hadron colliders like the Tevatron
through the decay modes:
$\eta_b \to J/\psi+J/\psi$ \cite{Braaten:2000cm},
and the recently proposed $\eta_b \to D^*D^{(*)}$ 
\cite{Maltoni:2004hv}, as the respective branching fractions would 
drop by about one order of magnitude
with respect to the SM calculations

\section{A MSSM scenario with CP violation}

It has been pointed out in the literature 
\cite{Pilaftsis:1998dd,Carena:2002bb,Godbole:2004xe} that 
CP violation in the Higgs sector of the MSSM can
occur quite naturally, representing an interesting option
to generate CP violation beyond the SM. Then, 
the three neutral MSSM Higgs bosons could mix together and
the resulting three physical mass eigenstates $H_1,H_2,H_3$ ($M_{H_1} < 
M_{H_2} < M_{H_3}$) would
have mixed parities. Under this scenario, Higgs 
couplings to the $Z$ boson would vary;
the $H_1ZZ$ coupling can be significantly suppressed \cite{Gunion:1997aq}
thus raising the possibility of a relatively light $H_1$ 
boson having escaped detection at LEP 2 
for the range $10\ {\lesssim}\ \tan{\beta}\ {\lesssim}\ 40$ 
(higher values of $\tan{\beta}$ were not considered in the search). 
Interestingly, for several choices of model parameters 
and a Higgs mass about 10 GeV, the region of
$\tan{\beta}$ not excluded
experimentally by LEP searches \cite{Abbiendi:2004ww} is  
in accordance with the requirements
found in this work to give rise to 
a $\sim 10\%$ breakdown of lepton universality
in $\Upsilon$ decays.

Finally notice that the $H_1$ Higgs boson can couple both to
scalar and pseudocalar states. Hence the
$\chi_{b0}$ resonances might 
play a role as intermediate states in $\Upsilon(2S)$ and
$\Upsilon(3S)$ decays: $\Upsilon \to \gamma_s \chi_{b0}(\to H_1 \to
\tau^+\tau^-)$, as mentioned in section 2.3 for a CP-even
Higgs boson.

\section{Summary}

Heavy quarkonium physics has reached a level of maturity enabling the
search for new phenomena beyond the SM \cite{Brambilla:2004wf}. 
In this paper, possible
hints of new physics in bottomonium systems and suggestions to
conduct an experimental search for new evidence
have been pointed out:
\begin{itemize}

\item[a)] Current experimental data do not preclude the possibility of 
lepton universality breaking 
at a significance level of $10\%$ \cite{Sanchis-Lozano:2003ha}, 
interpreted in terms of a light CP-odd Higgs boson for 
a reasonable range of $\tan{\beta}$ values in a 2HDM(II).
In fact, direct searches at LEP don't exclude a 
non-standard Higgs boson for some 
regions of model parameters in different scenarios, 
notably in a CPX MSSM \cite{Abbiendi:2004ww}
and the NMSSM \cite{Hiller:2004ii}

\item[b)] Mixing between the CP-odd Higgs and $\eta_b$ states can 
yield $M_{\Upsilon(nS)}-M_{\eta_b(nS)}$ splittings larger 
than expected within the SM if $M_{A_0^0} > M_{\eta_{b0}}$;
the opposite if $M_{A_0^0} < M_{\eta_{b0}}$. Furthermore,
large $\eta_b$ widths would also be  
expected for large $\tan{\beta}$ values. All
that might explain the failure to find any signal from 
hindered $\Upsilon(2S)$ and $\Upsilon(3S)$ magnetic dipole transitions
into $\eta_b$ states despite intensive searches performed at CLEO
\cite{Artuso:2004fp,Blusk:2004mm}. 
Negative effects would also come up in the prospects to
detect $\eta_b$ resonances in hadron colliders like the Tevatron

\item[c)]  After the recent results by CLEO 
\cite{Danko:2004sb,Blusk:2004mm,Adams:2004xa} on the 
muonic BF of all three $\Upsilon(1S)$,
$\Upsilon(2S)$, $\Upsilon(3S)$ resonances, new results for the tauonic
BF from on-going CLEO analysis are 
eagerly awaited. Indeed,  
distinct degrees of lepton universality breaking (if any) in
Upsilon decays 
might lead to a Higgs mass estimate as one 
expects that the closer the resonance mass to the mediating Higgs 
boson mass is, the bigger NP effect shows up in the tauonic BF

\item[d)] The detection of (quasi)monoenergetic photons (assuming a sharp 
intermediate $\eta_b$ state) in $\sim 10\%$ events of the
tauonic decay sample would represent the $\lq\lq$smoking gun'' 
of a Higgs boson mediating the decay, allowing the determination of
its mass. Moreover, 
polarization studies of final-state $\tau^{\pm}$ in
$\Upsilon$ decays \cite{Kramer:1993jn} could help to establish
also the CP quantum numbers of the Higgs boson

\item[e)] On the contrary, if lepton universality in $\Upsilon$ decays
were confirmed, mass windows still open for a light Higgs boson 
in scenarios beyond the SM could be closed
in parameter regions hardly reachable by other experiments, e.g. those
performed at hadron colliders 
\cite{Ellwanger:2003jt,Gunion:2004si}. This would be 
especially the case for a Higgs particle
below open bottom production

\end{itemize}

\subsection*{Acknowledgements}
I gratefully acknowledge S. Baranov, N. Brambilla, 
J.F. Gunion, A. Vairo, the
Quarkonium Working Group and 
the Analysis Group of the CLEO Collaboration for 
useful comments and discussions.

\end{document}